\documentclass[12pt,preprint]{aastex}
\def\teff{\ensuremath{T_{\rm eff}}}

\def\ltsima{$\; \buildrel < \over \sim \;$}
\def\gtsima{$\; \buildrel > \over \sim \;$}
\def\lsim{\lower.5ex\hbox{\ltsima}}
\def\gsim{\lower.5ex\hbox{\gtsima}}
\def\lapp{\ifmmode\stackrel{<}{_{\sim}}\else$\stackrel{<}{_{\sim}}$\fi}
\def\gapp{\ifmmode\stackrel{>}{_{\sim}}\else$\stackrel{<}{_{\sim}}$\fi}
\newdimen\minuswidth    
\setbox0=\hbox{$-$}
\minuswidth=\wd0
\catcode`@=\active
\def@{\kern\minuswidth}
\setbox0=\hbox{\rm0}
 
\shorttitle{Blue Stragglers in NGC6388}
\shortauthors{Dalessandro et al.}
 
\begin{document} 
\title{Blue Straggler Stars in the Unusual Globular Cluster NGC~6388\footnote{Based on observations with the NASA/ESA {\it HST}, obtained at the
Space Telescope Science Institute, which is operated by AURA, Inc., under
NASA contract NAS5-26555. Also based on WFI observations collected
at the European Southern Observatory, La Silla, Chile.}}

\author{
E. Dalessandro\altaffilmark{1,2},
B. Lanzoni\altaffilmark{1},
F.R. Ferraro\altaffilmark{1},
R.T. Rood\altaffilmark{3},
A. Milone\altaffilmark{4},
G. Piotto\altaffilmark{4}.
E. Valenti\altaffilmark{5}
}
\affil{\altaffilmark{1} Dipartimento di Astronomia, Universit\`a degli Studi
di Bologna, via Ranzani 1, I--40127 Bologna, Italy}
\affil{\altaffilmark{2} ASI,Centro di Geodesia Spaziale, contrada Terlecchia,
I-75100, Matera, Italy }   
\affil{\altaffilmark{3} Astronomy Department, University of Virginia,
P.O. Box 400325, Charlottesville, VA, 22904}
\affil{\altaffilmark{4} Dipartimento di Astronomia, Universit\`a degli Studi
di Padova, vic. Osservatorio 2, I--35122 Padova, Italy}
\affil{\altaffilmark{5} European Southern Observatory,
 Alonso de Cordova 3107, Vitacura, Santiago, Chile}
\date{2 October, 07}

\begin{abstract}

We have used multi-band high resolution {\it HST} WFPC2 and ACS
observations combined with wide field ground-based observations to
study the blue straggler star (BSS) population in the galactic
globular cluster NGC~6388.  As in several other clusters we have
studied, the BSS distribution is found to be bimodal: highly peaked in
the cluster center, rapidly decreasing at intermediate radii, and
rising again at larger radii. In other clusters the sparsely populated
intermediate-radius region (or ``zone of avoidance'') corresponds well
to that part of the cluster where dynamical friction would have caused
the more massive BSS or their binary progenitors to settle to the
cluster center.  Instead, in NGC~6388, BSS still populate a region
that should have been cleaned out by dynamical friction effects, thus
suggesting that dynamical friction is somehow less efficient than
expected.  As by-product of these observations, the peculiar
morphology of the horizontal branch (HB) is also confirmed. In
particular, within the (very extended) blue portion of the HB we are
able to clearly characterize three sub-populations: ordinary blue HB
stars, extreme HB stars, and blue hook stars. Each of these
populations has a radial distribution which is indistinguishable from
normal cluster stars.

\end{abstract}
 

\section{INTRODUCTION}

The present work is a continuation of a series of papers exploring the
interplay of dynamics and stellar evolution in Galactic globular clusters
(GCs), through the detailed study of their peculiar stellar populations.  To
date we have primarily focused on Blue Straggler Stars (BSS). BSS are
brighter and hotter than the Main Sequence (MS) Turnoff (TO) and are thought
to form by the evolution of primordial binaries and/or by the effect of
stellar dynamical interactions. They are most abundant in the central regions
of GGCs, which are usually very congested in the optical bands, but
relatively open in the UV. To fully understand the role of dynamics one must
track how the population of BSS (or any other population arising from binary
evolution or stellar interactions) varies over the entire radial extent of
the cluster. The results of the earlier papers in the series are summarized
in Table~\ref{tab:papers}. 

Of the clusters we have explored so far NGC~6388 is perhaps the strangest. As
shown for the first time by Rich et al. (1997), this bulge cluster has an
extended blue Horizontal-Branch (HB), despite its high metal content
\citep[${\rm [Fe/H]}=-0.44$;][]{carretta07}. Such an HB morphology is
contrary to expectations: metal-poor clusters usually have predominantly blue
HBs, while metal-rich clusters have stubby red HBs. NGC~6388 is a vivid
example of the so-called second-parameter phenomenon \citep[see,
e.g.,][]{swcat98, catelan06}.  Moreover, hints for the presence of an
intemediate mass black hole (IMBH) at the center of NGC~6388 have been
recently inferred from the observed surface brightness profile (which shows a
deviation from a flat core behavior in the inner $\sim 1\arcsec$) and from
detailed dynamical modeling \citep{miocchi07, lan07bh}.  More robust
evidence (through kinematics, X-ray and/or radio observations) is
needed to confirm this finding.

In this paper we present multi-wavelength observations of NGC~6388. We have
combined high-resolution ultraviolet (UV) and optical images of the cluster
center obtained with {\it HST}, with complementary wide-field observations
from the ground, covering the entire cluster extension. The data and
photometric reductions are described in \S~\ref{sec:data}.  A general
overview of the color-magnitude diagram (CMD) is discussed in
\S~\ref{sec:cmd}. The BSS population is described in \S~\ref{sec:BSS}, and
the Discussion is presented in \S~\ref{sec:discus}.

\section{OBSERVATIONS AND DATA ANALYSIS}
\label{sec:data}

\subsection{The data sets}
We have used a combination of high-resolution and wide-field photometric data
sets: 

{\it 1. The High resolution set} consists of a series of public multiband
(from the UV, to the optical) WFPC2 and ACS {\it HST} images, which have been
retrieved from the ESO/ST-ECF Science Archive.  The WFPC2 images were
obtained through filters $F255W$ (mid-UV) and $F336W$ ($U$), with total
exposure times ($t_{\rm exp}$) of 9200 and 1060 s, respectively (Prop. 8718,
P.I. Piotto), and trough filters $F439W$ ($B$) and $F555W$ ($V$), with
$t_{\rm exp}= 370$ and 62 s, respectively (Prop. 6095, P.I. Djorgovski).
This combined dataset allows us to examine both the hot (HB and BSS) and the
cool (red giant branch and subgiant branch, RGB and SGB, hereafter) stellar
populations in the cluster.  In this dataset the planetary camera (PC, with
the highest resolution of $\sim 0\farcs{046}\,{\rm pixel}^{-1}$) was roughly
centered on the cluster center, while the wide field cameras (WFCs, at a
lower resolution of $\sim 0\farcs{1}\,{\rm pixel}^{-1}$) sampled the
surrounding outer regions.  The photometric reduction of these images was
performed using ROMAFOT (Buonanno at al. 1983), a package developed to
perform accurate photometry in crowded regions and specifically optimized to
handle under-sampled Point Spread Functions (PSFs; Buonanno \& Iannicola
1989), as in the case of the WF chips.
The ACS dataset is composed of a series of images (Prop. 9821, P.I. Pritzl)
obtained through filters $F435W$ ($B$) and $F606W$ ($V$), with $t_{\rm
exp}=11$ and 7\,s, respectively.  It gives complete coverage of the central
cluster region out to $110\arcsec$ from the center (see the following
section).  All the ACS images were corrected for geometric distortions and
effective flux (Sirianni et al. 2005). The photometric analysis was performed
independently in the two images by using the the aperture photometry code
SExtractor (Bertin \& Arnouts 1996) and adopting an aperture radius of
3~pixels (corresponding to 0\farcs15).
 
{\it 2. The Wide field set} is a complementary set of public $B$ and $V$
images obtained with the Wide Field Imager (WFI) mounted at the 2.2m ESO-MPI
telescope at La Silla ESO-Observatory and retrieved from the ESO
Science Archive. The WFI is a mosaic of 8-CCD chips, each of $2000\times
4000$ pixels, with a pixel size of $\sim 0\farcs24$. It has exceptional
image capability, with a global field of view (FoV) of $33\arcmin\times
34\arcmin$.  Thanks to such a FoV, this dataset covers the entire cluster
extension with the cluster roughly centered on the CCD \#7.  The WFI images
have been corrected for flatness, bias and overscan using IRAF tools. The PSF
fitting was performed independently on each image using DAOPHOT \citep{dao}.

\subsection{Astrometry, Photometric Calibration, and Sample Definition}
Using the procedure described in Ferraro et al. (2001, 2003) the WFI
catalogue has been placed on the absolute astrometric system.  The 8 WFI CCDs
have been astrometrized by cross-correlating each of them with the new
astrometric 2-MASS catalogue using a specific tool developed at Bologna
Observatory.  Several hundred astrometric reference stars were found in each
WFI chip, thus allowing an accurate absolute positioning of the sources.  As
a second step, a few hundred stars in the overlapping area between the WFI,
and the WFPC2 and ACS FoVs were used as secondary astrometric standards, in
order to place the {\it HST} catalogs on the absolute astrometric system. At
the end of the procedure the rms residuals (that we take as representative of
the astrometric accuracy) were of the order of $\sim 0\farcs 3$ both in RA
and Dec.

By using the procedure described in Ferraro et al. (1997, 2001), the
photometric calibration of the UV magnitudes ($F255W$ and $F336W$) has
been performed in the STMAG system, adopting the \citet{holtz95}
zero-points. The optical ($B$ and $V$) magnitudes have been
transformed to the Johnson system by using the stars in common with
the catalog of \citet{pio02}.  Linear transformations were adopted,
and only small color equation terms were required to correct the
response of the different filter profiles.

\begin{figure}[!hp]
\begin{center}
\includegraphics[scale=0.7]{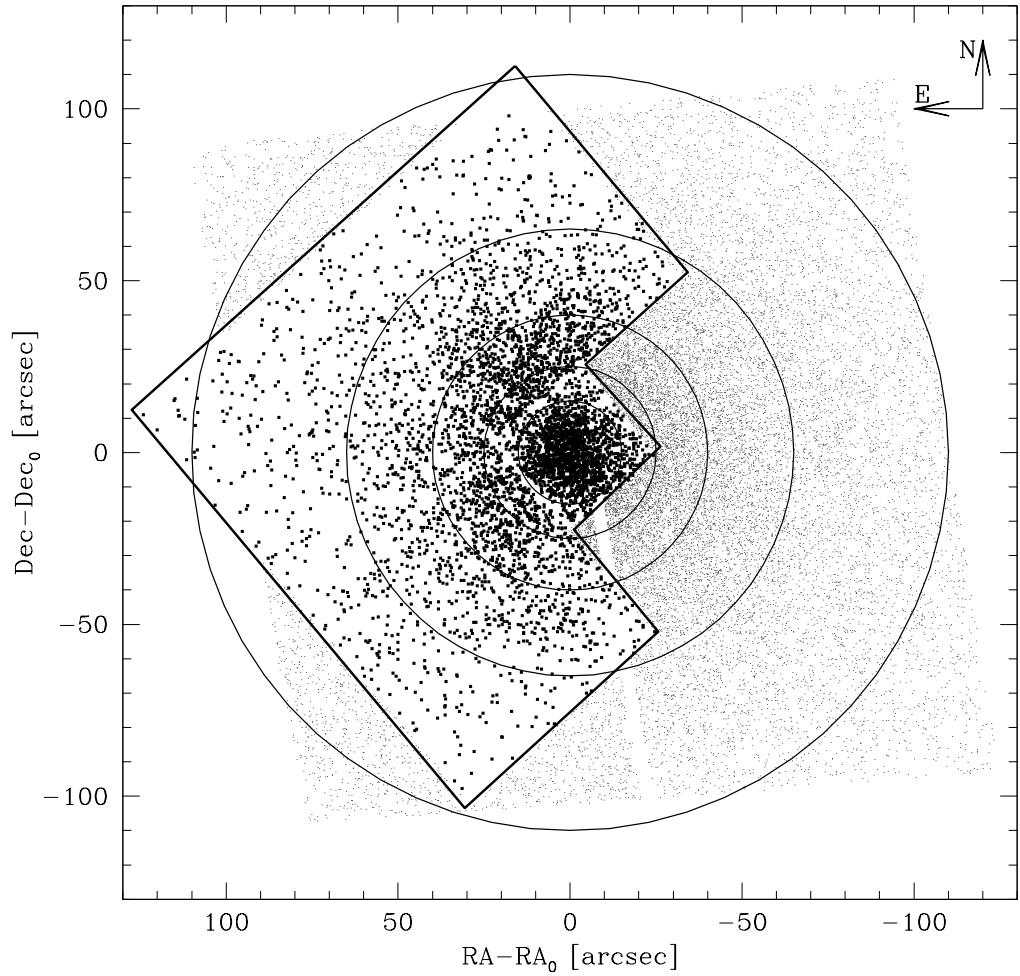}
\caption{Map of the {\it HST sample}. The thick solid line delimits the WFPC2
{\it HST} FoV. The concentric annuli are used to study the radial
distribution of BSSs. The inner and the outer annuli correspond to
$r=5\arcsec$ and $r=110\arcsec$, respectively.}
\label{fig:maphst}
\end{center}
\end{figure}

\begin{center}
\begin{figure}[!p]
\includegraphics[scale=0.7]{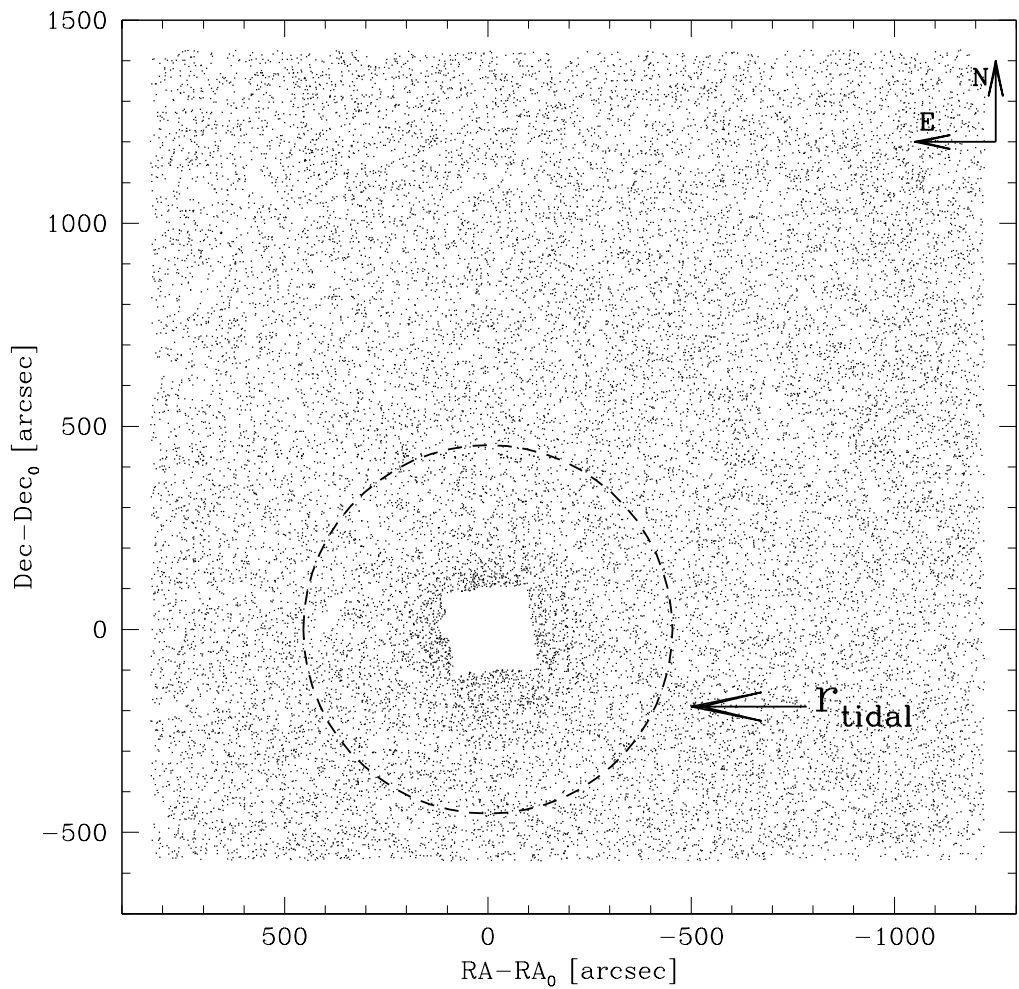}
\caption{Map of the {\it WFI sample}. This sample has been used to estimate
the structural parameters of NGC~6388 and the Galaxy contamination, but not
for constructing the radial distribution of BSS. The dashed line marks the
cluster tidal radius ($r_t=454\arcsec$).}
\label{fig:mapwfi}
\end{figure}
\end{center}

Final lists with the absolute coordinates and homogeneous magnitudes for all
the stars in the three considered catalogs were obtained.  To minimize
incompleteness effects in the ground based observations of the crowded
central regions of the cluster, while still taking advantage of the superior
capability of UV observations in detecting the BSS (Ferraro et al 1999a,
2001), we divided the dataset in two main samples: The {\it HST sample},
includes only stars in the WFPC2 and complementary ACS catalogs. It covers
approximately the inner $r<110\arcsec$ of the system (with the WFPC2 FoV
almost entirely included in the ACS FoV; see Figure~\ref{fig:maphst}). The
{\it WFI sample} includes only stars observed with WFI and lying beyond the
WFPC2 and ACS FoVs.  The {\it WFI sample} covers the outer regions, well
beyond the cluster extension (see Figure~\ref{fig:mapwfi}).

By combining the data sets described above, with additional images of the
cluster center obtained with the {\it HST} ACS High Resolution Camera, very
accurate determinations of the center of gravity, surface density profile,
and surface brightness profile have been recently obtained by
\citet{lan07bh}. In particular, it has been found that the observed profiles
show a deviation from a flat core behavior in the inner $\sim 1\arcsec$,
suggesting that NGC~6388 might host an IMBH of $\sim 5.7\times 10^3\,
M_\odot$ in its center.  However, by excluding the points at $r<1\arcsec$,
the density profile is well fit by an isotropic single-mass King model. The
resulting cluster structural parameters (concentration, core radius and tidal
radius) are listed in Table~\ref{tab:param}, together with the new estimate
of the center of gravity.

\section{CMD OVERVIEW}
\label{sec:cmd}
\subsection{The HST Sample}
\label{sec:hst}
\begin{figure}[!p]
\begin{center}
\includegraphics[scale=0.7]{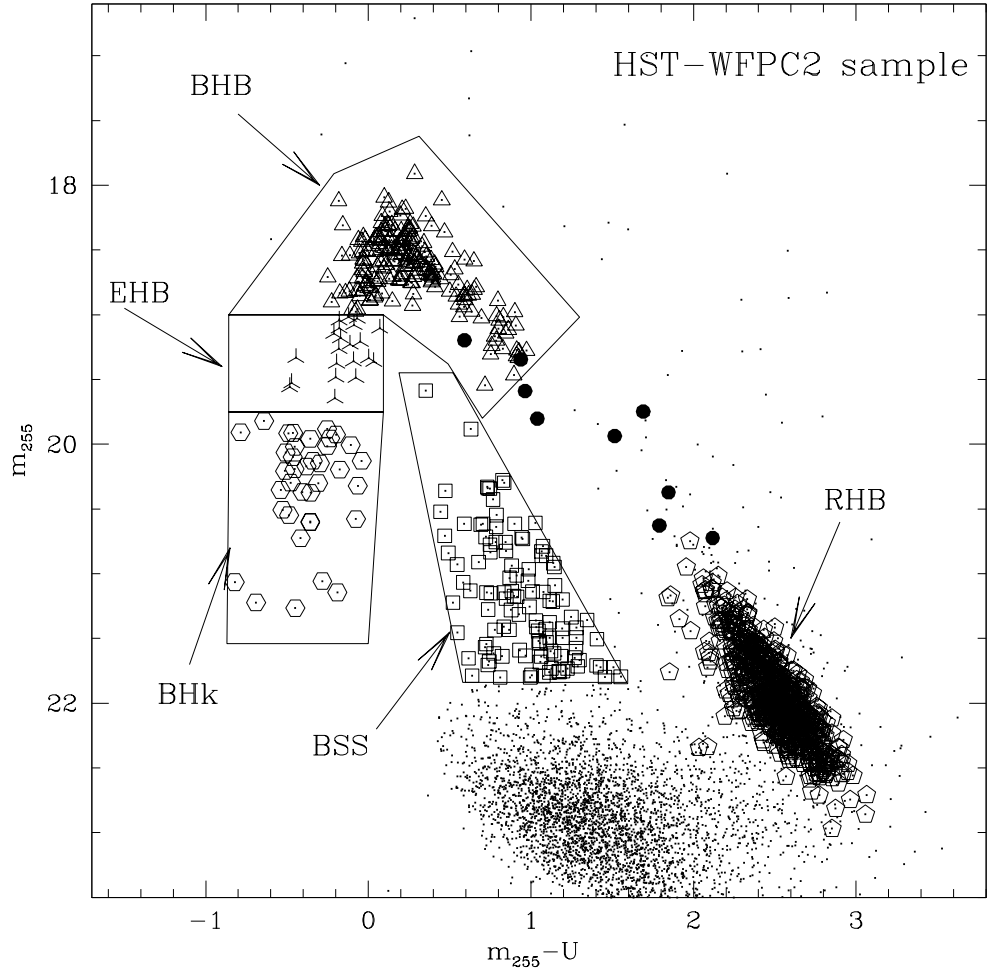}
\caption{Ultraviolet CMD of the {\it HST} WFPC2 sample. The different stellar
populations discussed in the paper are marked with different symbols, as
indicated by the labels. Solid dots mark the selected RR Lyrae stars.}
\label{fig:uvcmd}
\end{center}
\end{figure}

\begin{figure}[!p]
\begin{center}
\includegraphics[scale=0.7]{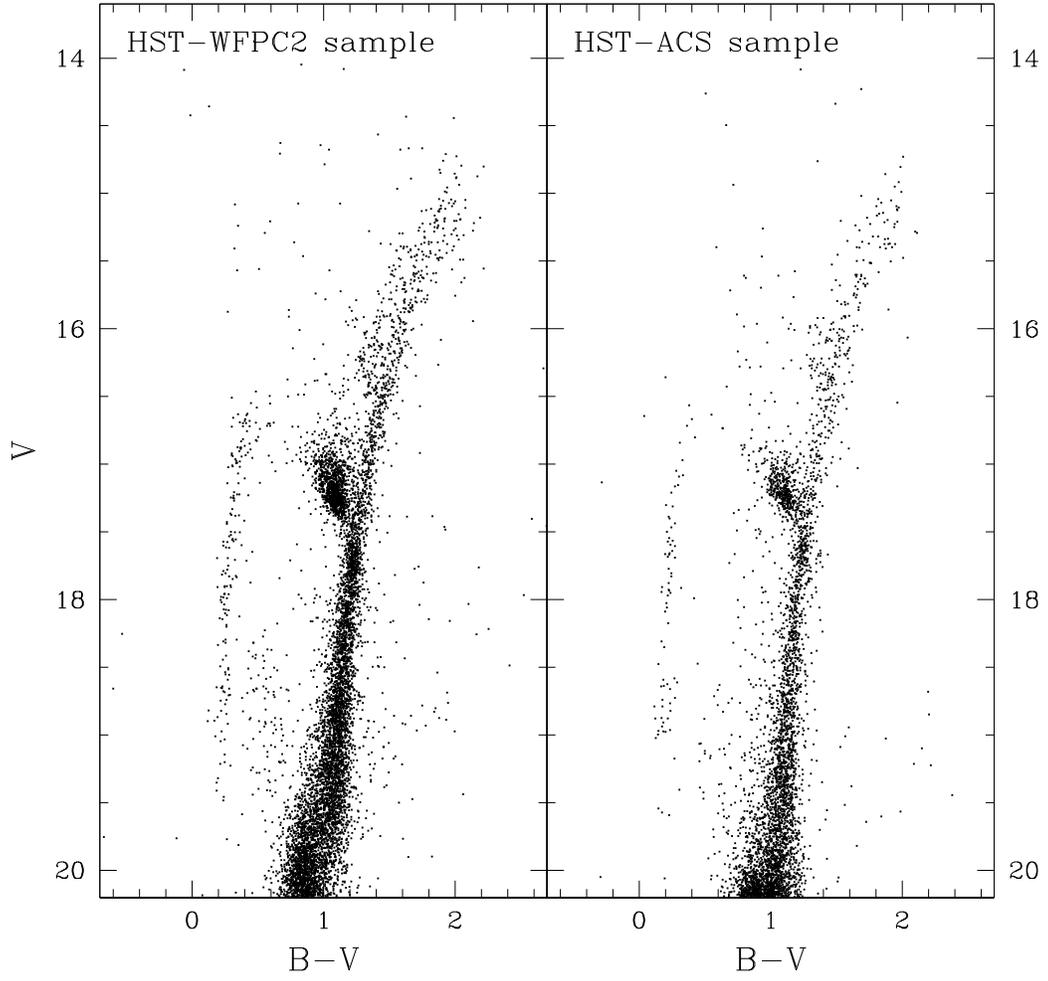}
\caption{Optical CMDs of the {\it HST} WFPC2 and the complementary ACS
samples.}
\label{fig:hstcmd}
\end{center}
\end{figure}

The CMDs of the {\it HST sample} in the UV and the
optical bands are shown in Figures~\ref{fig:uvcmd} and \ref{fig:hstcmd},
respectively.  As apparent, all of the cluster evolutionary sequences are
clearly defined and well populated.

Particularly notable is the Horizontal Branch (HB) morphology.  Beside the
red clump, which is a typical feature of metal rich stellar populations, the
HB clearly shows an extended blue tail (BT), first noticed by Rich et
al. (1997) and by Piotto at al. (1997).  Among a total of 1763 HB stars
counted in the {\it HST sample}, five sub-populations can be distinguished
(see Sect.~\ref{sec:HB} and \ref{sec:RHB} for details): {\it (i)} the red-HB
(RHB) population, consisting of 1418 stars grouped in the red clump; {\it
(ii)} 15 RR Lyrae variables, which we identified by cross correlating the
positions in our catalog with those published by Pritzl et al. (2002)
\footnote{Since our photometry is just a snapshot, the position of each
variable star in the CMD is not an indication of the mean properties.  The
remaining stars found within the "RR Lyrae region" of the CMD, but not
included in the Pritzl et al. catalog are not considered in the following
analysis, since they possibly are field contaminating stars.}; {\it (iii)}
267 blue-HB (BHB) stars ; {\it (iv)} 26 Extreme-HB (EHB) stars ; and {\it
(v)} 37 Blue Hook (BHk, to avoid confusion with BH for black hole) stars.

Several previous works have shown that the HB morphology in NGC~6388 is
complex. A new extensive study, based on much of the same observational data
used here, and also discussing the HB morphology of NGC~6441, has recently
been published by \citet{busso07}.  In the present paper we take advantage of
the complex HB structure of NGC~6388 to review the HB nomenclature, which has
become rather confused in the literature and is often ambiguously used. Then,
we briefly discuss the blue HB sub-populations and the HB red clump, the
latter being used as cluster reference population for the study of the BSS
radial distribution (Sect.~\ref{sec:radist}).

\subsubsection{The HB population: nomenclature and  radial distribution}
\label{sec:HB}
The HB is composed by helium-core/hydrogen-shell burning stars. It is
traditionally split into red, variable, and blue (RHB, VHB, and BHB,
respectively), depending on whether the stars are redder than, within, or
bluer than, the RR~Lyrae instability strip.

The concept of {\em HB blue tails} probably originated with the CMD of
NGC~6752, which was presented by Russell Cannon at the 1973 Frascati globular
cluster workshop, but not published for many years. In visual CMDs of
NGC~6752 (and many others to follow), the HB drops downward at high
temperature often becoming an almost vertical sequence. This feature looked
like a tail hanging from the horizontal part of the BHB, hence the
name. \citet{rc89}, \citet{bt-dens93}, and \citet{recio-blanco06} have each
suggested ways to measure BTs, and the fact that measures of BTness keep
being invented demonstrates a lack of consensus on a definition of BTs. In
addition, sub-populations like EHB or BHk stars are sometimes recognized
within the observed BTs, even if without a precise observational definition.

The {\em extreme HB} population is theoretically well defined: EHB stars lie
at the hottest extreme of the zero-age HB (ZAHB), and they do not return to
the asymptotic giant branch (AGB), but rather spend their He-shell burning
phase as hot AGB-manqu\'e or Post-early AGB stars \citep[for
example,][]{dor93}. There is no comparably precise way to observationally
select EHB stars. If far-UV (FUV) (e.g., {\it HST} F160W) photometry is
available, detailed comparisons with stellar models can be made. These
suggest that in a few clusters, the transition between BHB and EHB stars may
be associated with a gap in the HB morphology \citep{fe98}.  In the present
paper we have assumed this to be the case (see below), but at this point that
is only an assumption.  The importance of EHB stars is also connected with
the fact that they and their progeny are thought to be the source of the UV
radiation excess observed in the integrated spectra of some elliptical
galaxies (Dorman et al 1995; Yi et al. 1998), and one might be able, for
example, to determine the age of the galaxy on the basis of its UV excess. In
this context NGC~6388 plays a particularly important role, since it is one of
the most metal-rich systems containing EHB stars that can be individually
observed.

In a few clusters, including NGC~6388, there is an additional population
hotter and less luminous than the EHB stars. Following nomenclature used in
recent studies, we call this population {\em Blue Hook} stars. In visual and
even some UV (e.g., $m_{255},~m_{255}-m_{336}$) CMDs, BHk stars appear as
fainter extension of the BT and are separated from the EHB population by a
gap.  While the effective temperature \teff\ of HB stars can be reasonably
well determined from their position along the BT, it is not appropriate to
extrapolate this to the BHk. Accurate stellar parameters for BHk stars
require FUV photometry \citep[see for example the BHk studies in NGC~2808 and
$\omega$\,Cen by][]{moehler2808}. Indeed, it is only in FUV CMDs that the
origin of the name ``blue hook'' becomes apparent.

\begin{figure}[!p]
\begin{center}
\includegraphics[scale=0.7]{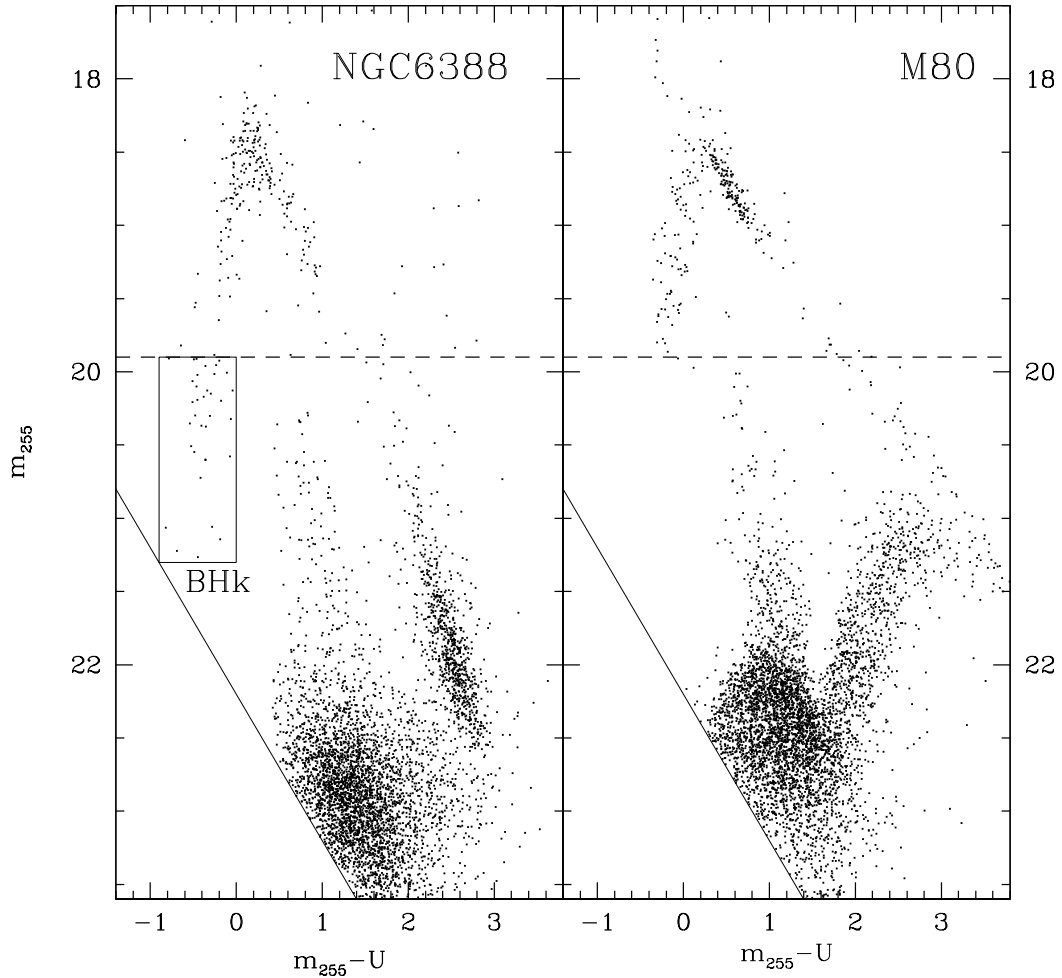}
\caption{Comparison between the UV CMDs of NGC~6388 and M80. The latter
has been suitably shifted in color and magnitude in order to superimpose the
knees of the two HBs (at $m_{255}-U\simeq 0.2$ and $m_{255}\simeq 18.5$). The
dotted line marks the limit ($m_{255}\simeq 20$) below which there are no
more BHB stars in M80, and that we have adopted as the brightest boundary of
the BHk population.}
\label{fig:m80}
\end{center}
\end{figure}

\begin{figure}[!p]
\begin{center}
\includegraphics[scale=0.7]{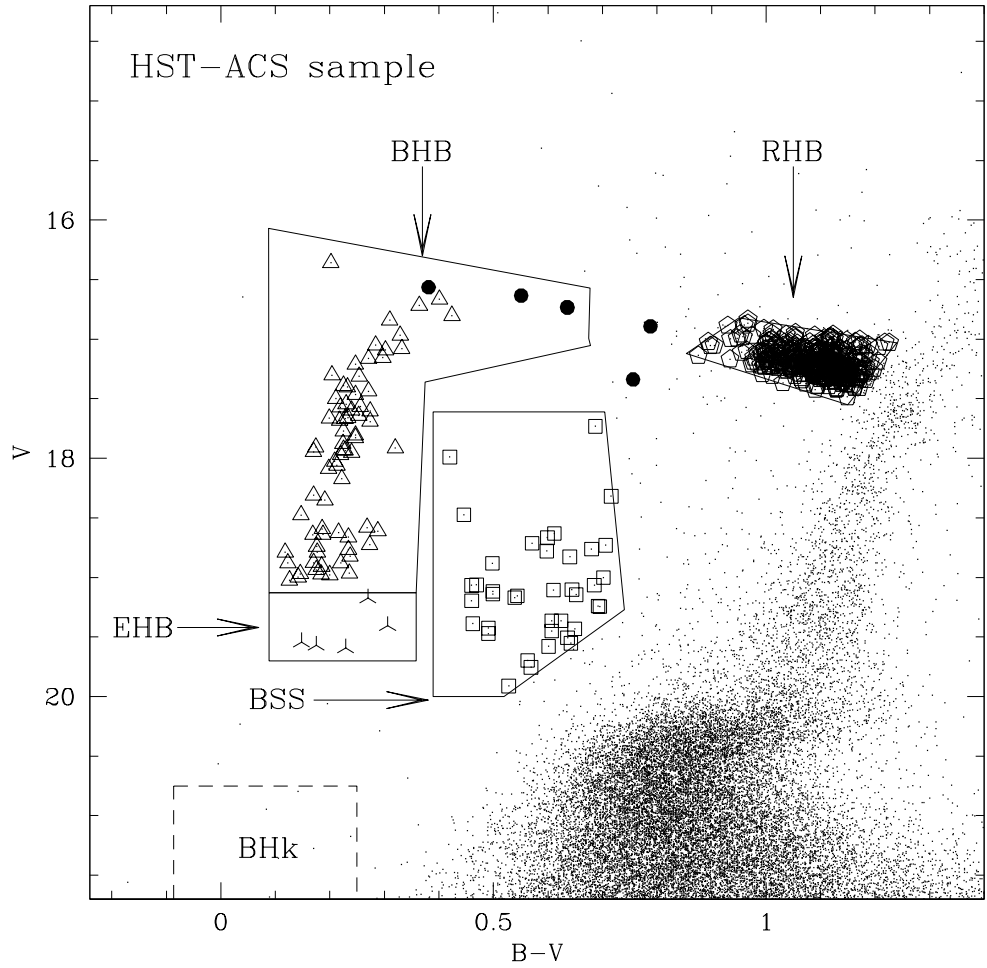}
\caption{Zoom of the optical CMD of the complementary {\it HST} ACS
sample, showing the different stellar populations discussed in the paper,
marked as in Fig.~\ref{fig:uvcmd}.  The selection box of BHk stars is marked
as a dashed line, since this population is near the detection limits in the
optical wavelengths.}
\label{fig:Optsel}
\end{center}
\end{figure}

Not all BT clusters have EHB stars \citep[see the case of
NGC~1904][]{lan07_1904}, and not all clusters with EHB stars have BHk stars
\citep[see the cases of M13 and M80][]{fe98}. In order to clearly show the
difference between cluster with BTs populated up to the EHB region, and
clusters with BHk stars, UV photometry is essential.  In Figure~\ref{fig:m80}
we compare the ($m_{255},~m_{255}-U$) CMDs of NGC~6388 to that of M80
(Ferraro et al. 1997, 1999a).  We chose M80 because, among the clusters for
which we have a full range of data, it is the one with the HB extending to
highest \teff.  From comparison with evolutionary tracks in the
($m_{160},~m_{160}-V$) plane (Dorman and Rood, unpublished) we know that the
M80 HB is populated all the way to the extreme blue-end of the ZAHB.  In this
plane the HB of M80 shows a clear gap at the transition from BHB to EHB
\citep{fe98}. In Fig.~\ref{fig:m80} this gap is also visible at $m_{255} =
19$. Since there is a corresponding gap in NGC~6388, we tentatively identify
the stars with $19.0 < m_{255} <19.8$ as EHB.  The HB sequence of NGC~6388 is
significantly more extended than that observed in M80, where there is no
analogous population at $m_{255} \simeq 20$. For that reason we identify the
latter as BHk stars in NGC~6388, as do \citet{busso07}. The BHk extends well
below the cluster TO in the optical CMD: $V>21$ in the $(V,~B-V)$ plane (see
Fig.~\ref{fig:Optsel}).  Because of the uncertainties in detection of these
faint stars using optical wavelengths in such crowded regions, our BHk sample
is selected from the UV CMD (WFPC2 sample).  By using these criteria we have
defined the selection boxes sketched in Figs.~\ref{fig:uvcmd} and
\ref{fig:Optsel}, and obtained the values quoted in the previous section for
the number of HB stars belonging to each sub-population.

\begin{figure}[!p]
\begin{center}
\includegraphics[scale=0.7]{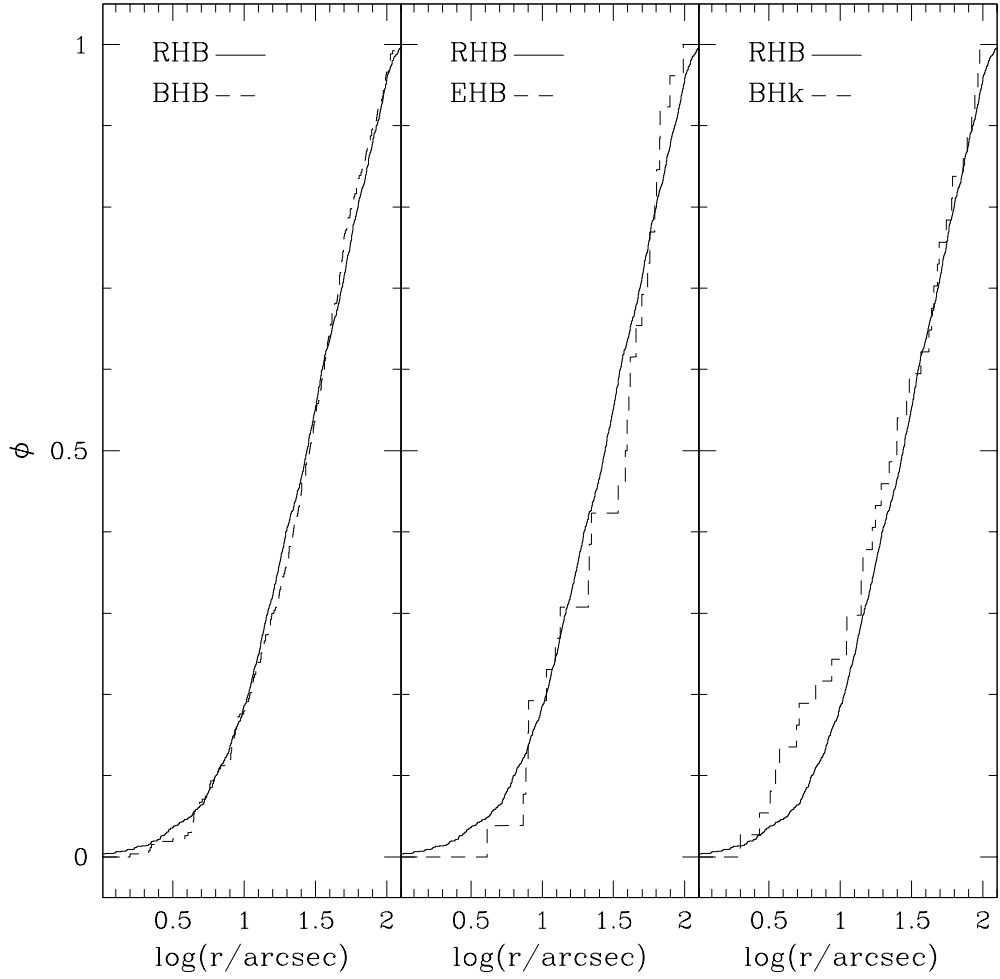}
\caption{Cumulative radial distributions of BHB, EHB and BHk stars (dashed
lines), compared to that of the reference RHB population. No evidences of
peculiar radial distributions are found for the blue HB sub-populations, with
respect to normal cluster stars.}
\label{fig:HB_KS}
\end{center}
\end{figure}

It has been suggested that EHB might originate in binary systems
(Bailyn 1995), or they were formed through a collisional
channel. Indeed, similar stars in the field and in old open clusters
have been found to often belong to binary systems
\citep{green01,heber02}. In contrast, a recent study by Moni Bidin et
al. (2006) has found no evidence of binarity in the EHB population of
two globular clusters (M80 and NGC5986).  The nature of BHk stars is
still a matter of debate. They may be related to the so-called late
hot flashers \citep{moehler2808}, or to high helium abundances
\citep{busso07}.  Given that the origin of EHB and BHk stars is still
uncertain, it is useful to check whether their radial distributions
show any suggestion of binarity or stellar interaction, as it is the
case for BSS.  We have therefore compared the cumulative radial
distributions of the BHB, EHB, and BHk stars to that of RHB stars,
which are representative of normal cluster populations (see
Sect.~\ref{sec:radist}).  As shown in Figure~\ref{fig:HB_KS}, all
radial distributions are consistent with being extracted from the same
parent population, with Kolmogorov-Smirnov (KS) test probabilities of
59\%, 46\% and 60\%, respectively. Thus, the radial distribution of
BHB, EHB and BHk populations is consistent with that of normal cluster
stars, in agreement with previous findings by Rich et al. (1997), and
possibly suggesting a non-binary nature for these systems. However, a
binary EHB star could consist of a 0.5\,$M_\odot$ He-burning star with
a 0.2\,$M_\odot$ He white dwarf companion, i.e., a total mass smaller
than the TO mass and comparable to RHB masses. As a consequence, even
if the initial binary mass were large enough to have sunk to the
cluster core, the central relaxation time of NGC~6388 is $8.3 \times
10^7$\,yr \citep{djorg93}, less than the HB lifetime, so that an EHB
binary could move outward, instead of being segregated into the
center.

\subsubsection{The red HB clump and the distance of NGC~6388}
\label{sec:RHB}
Since the HB red clump in this cluster is very well defined in the optical
CMDs, we have selected the RHB population in this plane, and then used the
stars in common between the optical and the UV WFPC2 samples to identify it
in the $(m_{255},~m_{255}-U$) CMD. The selected stars are marked as
pentagons in Figs.~\ref{fig:uvcmd} and \ref{fig:Optsel}.

\begin{figure}[!p]
\begin{center}
\includegraphics[scale=0.7]{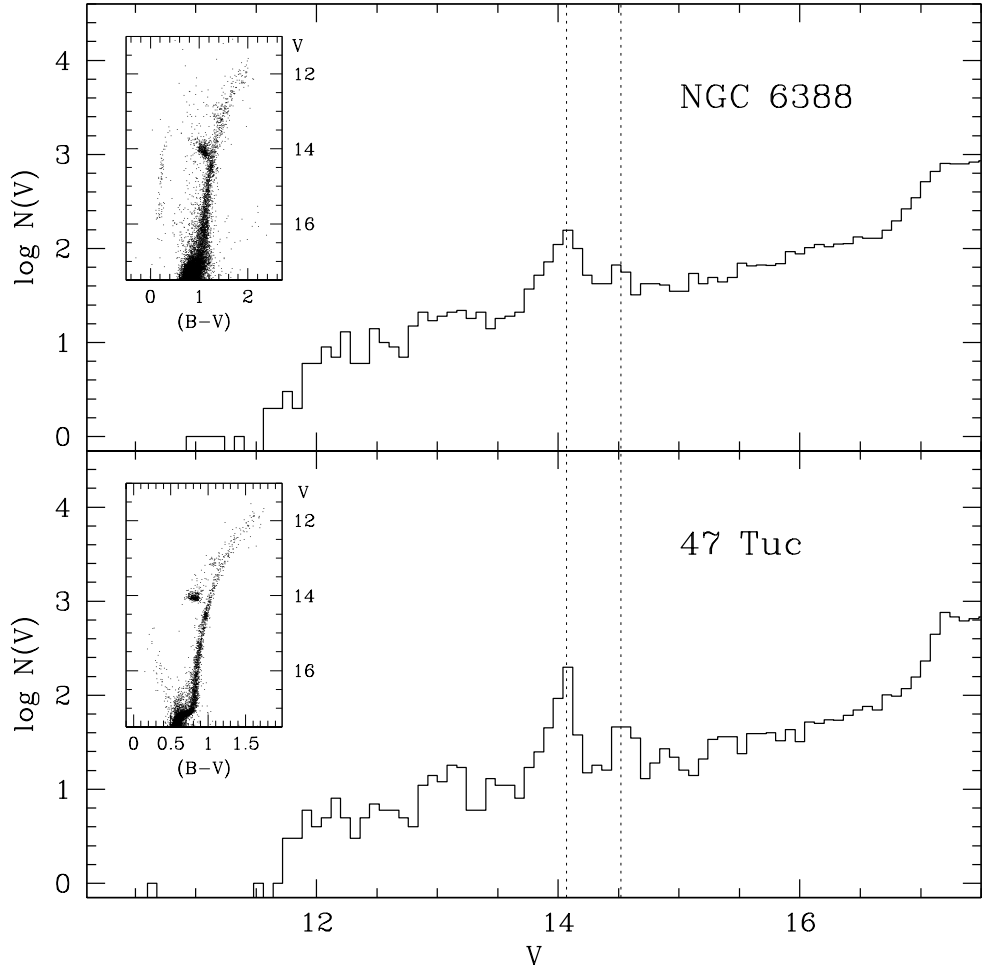}
\caption{Differential Luminosity Function of NGC~6388 shifted to that of
47~Tuc. The dotted lines indicate the HB red clump and the RGB-bump
level. From the inserts it is apparent that, other than the blue HB, the CMDs
of the two clusters are quite similar.}
\label{fig:47tuc}
\end{center}
\end{figure}

With our high-quality data set and such a well defined HB red clump,
we have estimated the distance modulus and the reddening of NGC~6388
by comparing its CMD and luminosity function to those of the
proto-type of metal-rich GCs: 47~Tuc.  As shown in
Figure~\ref{fig:47tuc}, other than the blue HB, the overall CMD
properties of NGC~6388 closely resemble those of a normal metal-rich
cluster.  In order to overlap the CMDs, and align the HB red clump and
the RGB bump of the two clusters, the color and the magnitude of
NGC~6388 have to be shifted by $\delta (B-V)=-0.28$ and $\delta
V=-3.15$, respectively. Thus, by adopting $(m-M)_0=13.32$ and
$E(B-V)=0.04$ for 47~Tuc (Ferraro et al. 1999b), we obtain
$(m-M)_V=16.59$ and $E(B-V)=0.32$ for NGC~6388. This yields a true
(unreddened) distance modulus of $ (m-M)_0=15.60 \pm 0.2$, which
corresponds to a distance of $13.2\pm 1.2$ Kpc \citep[to be compared
  with 10 Kpc quoted by][]{har96}.

The proper comparison of the CMDs of the two clusters deserves
additional comments.  First, NGC~6388 is known to be slightly more
metal-rich \citep[${\rm [Fe/H]}= -0.44$;][]{carretta07} than 47~Tuc
\citep[${\rm [Fe/H]} \simeq-0.6$;][]{carretta04}.  However, current
theoretical models suggest that such a small overabundance
($\delta{\rm [Fe/H]}\sim 0.15$ dex) would generate just a small
difference in the HB absolute magnitude ($\delta M_V^{\rm HB} \sim
0.03$ mag). Second, the presence of differential reddening of the
order of 0.07 (see Busso et al. 2007) can spread the HB red clump,
increasing the uncertainties of the entire procedure. However, this
contribution is significantly smaller than the conservative estimate
of the error in the derived cluster distance, $\delta(m-M)=\pm 0.2$
mag.  Finally, the morphology of the HB red clump is not exactly the
same in the two clusters. However, as discussed by \citet{catelan06},
who compared the two CMDs by using a reddening-independent quantity,
the main difference between the two red HB clumps consists in the fact
that the bluer RHB stars in NGC~6388 ($\sim 20\%$ of the total RHB
population) are slightly brighter than the average in 47~Tuc. This
feature might be interpreted in the framework of a sub-population with
a higher helium content. However comparison of the luminosity
functions in Figure~\ref{fig:47tuc} clearly shows that the relative
position of the HB red clump and RGB bump is quite similar in the two
clusters. Since the location in magnitude of the RGB bump is quite
sensitive to the helium content (see Fusi Pecci et al. 1990), such a
nice correspondence clearly demonstrates that at least the main
component of the stellar population of NGC~6388 has an helium
abundance fully compatible with that of 47~Tuc, while only a minor
fraction of the cluster stars could be helium enhanced (this is also
in agreement with the findings of Catelan et al. 2006, and Busso et
al. 2007). The possible impact on the relative distance of the two
clusters derived above is therefore negligible.

\subsection{The WFI Sample and Background Contamination}
\label{sec:contamination}
\begin{figure}[!p]
\begin{center}
\includegraphics[scale=0.7]{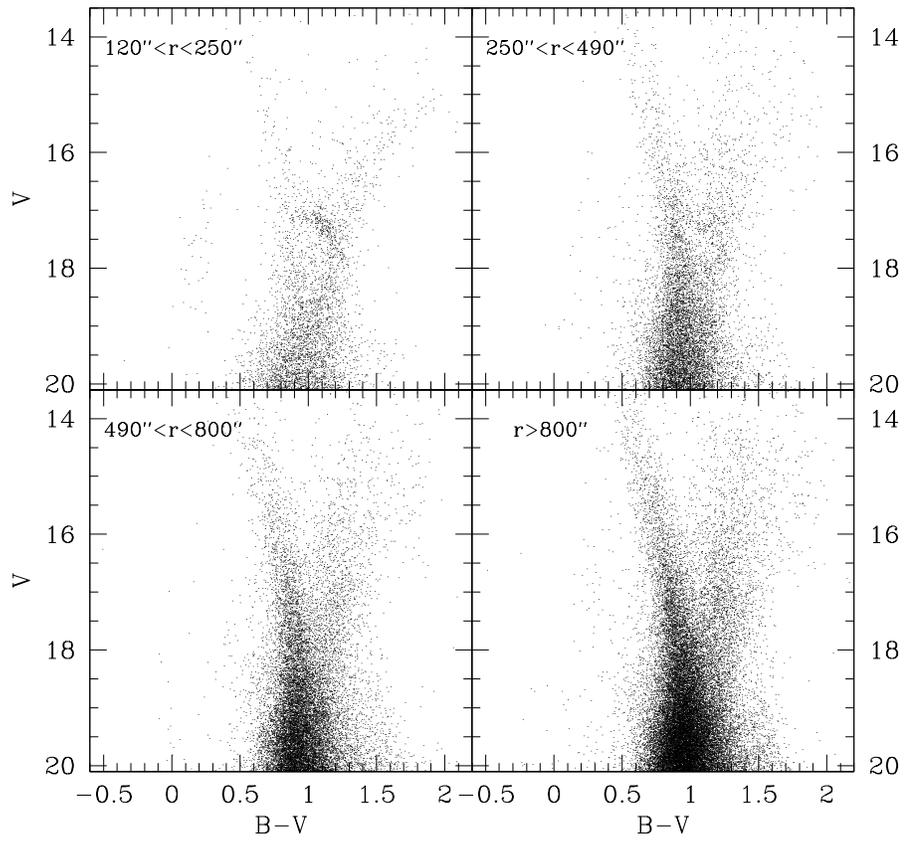}
\caption{Optical CMDs of the {\it WFI sample} for four different radial
ranges, as marked by the labels in each panel.}
\label{fig:wficmd}
\end{center}
\end{figure}

Figure~\ref{fig:wficmd} shows $(V,~B-V)$ CMDs of four radial zones of the WFI
sample. The sequences seen in Fig.~\ref{fig:hstcmd} are still obvious in the
interval $120\arcsec<r<250\arcsec$, but there is also significant
contamination from field stars.  The CMDs are progressively more contaminated
as $r$ increases: cluster sequences are barely visible in the region
$250\arcsec<r<490\arcsec$, less so for $490\arcsec<r<800\arcsec$, and have
vanished for $r>800\arcsec$. The contamination has two main components (the
bulge and the disk populations of the Galaxy): the first is an almost
vertical blue sequence with $0.5<(B-V)<1.0$, the second is another vertical
sequence with $(B-V)\sim 1.3$, which clearly indicates the presence of metal
rich stars.

Indeed, Figure~\ref{fig:wficmd} shows that field contamination is
particularly severe in this cluster. For this reason we decided to
limit the following analysis to the {\it HST sample}, and to use the
most external region of the WFI sample ($r > 800\arcsec$) to
statistically estimate the field contamination level.  We have counted
the number of background stars and derived an appropriate background
density for each selection box discussed in the paper (see
Sect.~\ref{sec:BSS} for the definition of the BSS population). Then,
we have used this background density to statistically decontaminate
each population in the {\it HST sample}, by following a procedure
similar to that described by \citet{bel99}. Table~\ref{tab:decon}
shows the observed sample size and the resulting statistical estimate
of field contamination for each of the sub-populations in each radial
region of the cluster.  Statistical decontamination has the
disadvantage that we don't know whether a given star is a member of a
given sub-population or not. However, all of the conclusions of the
current paper depend on number counts, so that background correction
only increases the noise without affecting the conclusions.  Future
proper motion studies currently ongoing for this cluster will finally
assess the real membership of each star.  Preliminary results indicate
contamination counts which are in agreement with those listed in Table
4.

\section{THE BLUE STRAGGLER STAR POPULATION}
\label{sec:BSS}
\subsection{The BSS Selection}
Hot populations like BSS and BHB stars are the
brightest objects in UV CMDs, while the RGB stars, that dominate the emission
of GCs in the optical bands, are faint at these wavelengths. In addition, the
high spatial resolution of {\it HST} minimizes problems associated with
photometric blends and crowding in the high density central regions. Thus
{\it HST} UV CMDs are the optimal tool for selecting BSS in GCs. Given this,
our main criterion for the selection of the BSS population is the position of
stars in the ($m_{255},~m_{255}-U$) CMD. To avoid incompleteness biases and
contamination from TO and sub-giant stars, we adopt a magnitude threshold
that is about one magnitude brighter than the TO point: $m_{255}= 21.85$.
Figure~\ref{fig:uvcmd} shows the adopted selection box and the candidate BSS
in the UV CMD.
Using the BSS in common between the WFPC2 and the ACS FoVs, we have
transformed the BSS selection box from the UV plane into the optical
plane. To avoid regions with very high risk of Galactic contamination, we
have considered only stars with $(B-V)<0.7$. The resulting candidate BSS in
the complementary ACS field are shown in Fig.~\ref{fig:Optsel}.

The final sample is of 153 BSS in the {\it HST sample}: 114 are found in the
WFPC2 dataset, and 39 in the complementary ACS sample.  The magnitude and the
positions of the selected BSS are listed in Table~\ref{tab:bss}, which is
available in full size in electronic form. 

\subsection{The BSS Projected Radial Distribution}
\label{sec:radist}
In order to study the radial distribution of BSS (or any other population)
for detecting possible peculiarities, a reference population representative
of normal cluster stars must be defined.  

In our previous papers we have used both the RGB or the HB as reference
populations.  In NGC~6388 the RGB population is affected by a significantly
larger field contamination, with respect to the HB. On the other hand, the HB
morphology is quite complex, the presence of a BT in such a metal-rich
cluster is unusual, and the nature of EHB and BHk stars is still unclear (see
Sect.~\ref{sec:HB}).  Instead, the HB red clump is a common feature of
similar metallicity GCs, it is bright and well defined both in the UV and in
the optical CMDs, and it comprises the majority (80\%) of the HB population
(see \S~\ref{sec:hst}).  We have verified that RHB and RGB stars
share the same radial distribution over the region ($r <
110\arcsec$), suggesting that RHB stars are indeed representative of normal
cluster populations.  For all these reasons, we have chosen the RHB as the
reference population for NGC~6388.

\begin{figure}[!p]
\begin{center}
\includegraphics[scale=0.7]{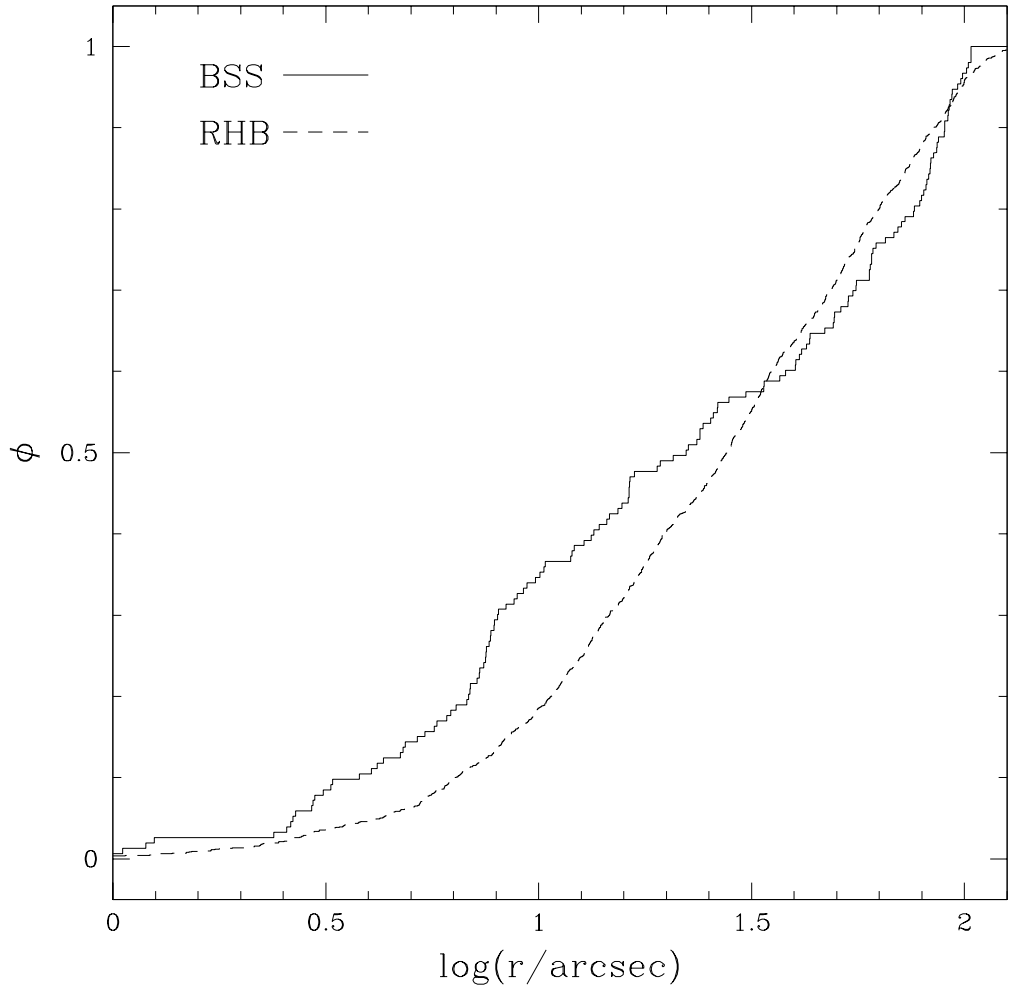}
\caption{Cumulative radial distribution of BSS (solid line) and RHB stars
(dashed line) as a function of the projected distance from the cluster
center, for the combined HST sample. The probability that they are extracted
from the same population is $\simeq 10^{-4}$.}
\label{fig:KS}
\end{center}
\end{figure}

We first compare the BSS and the RHB cumulative radial distributions.
Since we expect a negligible number of field stars contaminating the
BSS and RHB sample (see Table 4) no correction has been applied to the
observed sample used to construct these cumulative radial
distributions.  As shown in Figure~\ref{fig:KS}, the trend is bimodal,
with the BSS more segregated than RHB stars in the central cluster
regions, and less concentrated outward.  This result replicates what
we found in a number of other clusters and listed in Table~1 (See also
for M5 \citep{w06} and for M55 \citep{zag97}). The KS probability that the
two populations are extracted from the same parent distribution is
$\sim10^{-4}$.

For a more quantitative analysis we have divided the surveyed area in 6
concentric annuli (sketched in Fig.~\ref{fig:maphst}), and the number of BSS
($N_{\rm BSS}$) and RHB ($N_{\rm RHB}$) stars was counted in each annulus
(see the values listed in Table~\ref{tab:decon}). 
We have then computed the
double normalized ratio (Ferraro et al. 1993):
\begin{displaymath}
R_{\rm pop}=\frac{N_{\rm pop}/ N^{\rm tot}_{\rm pop}}{L^{\rm samp}/ L^{\rm
samp}_{\rm tot}},
\label{eq:Rpop}
\end{displaymath}
where pop=BSS, RHB, $N_{\rm pop}$ refers to statistically decontaminated
number counts (see Sect.~\ref{sec:contamination}), and the luminosity in each
annulus has been calculated by integrating the single-mass King model that
best fits the observed surface density profile \citep[see][]{lan07bh}, with
the distance modulus and the reddening previously quoted, and by properly
taking into account the incomplete spatial coverage of the outermost annulus.

\begin{figure}[!p]
\begin{center}
\includegraphics[scale=0.7]{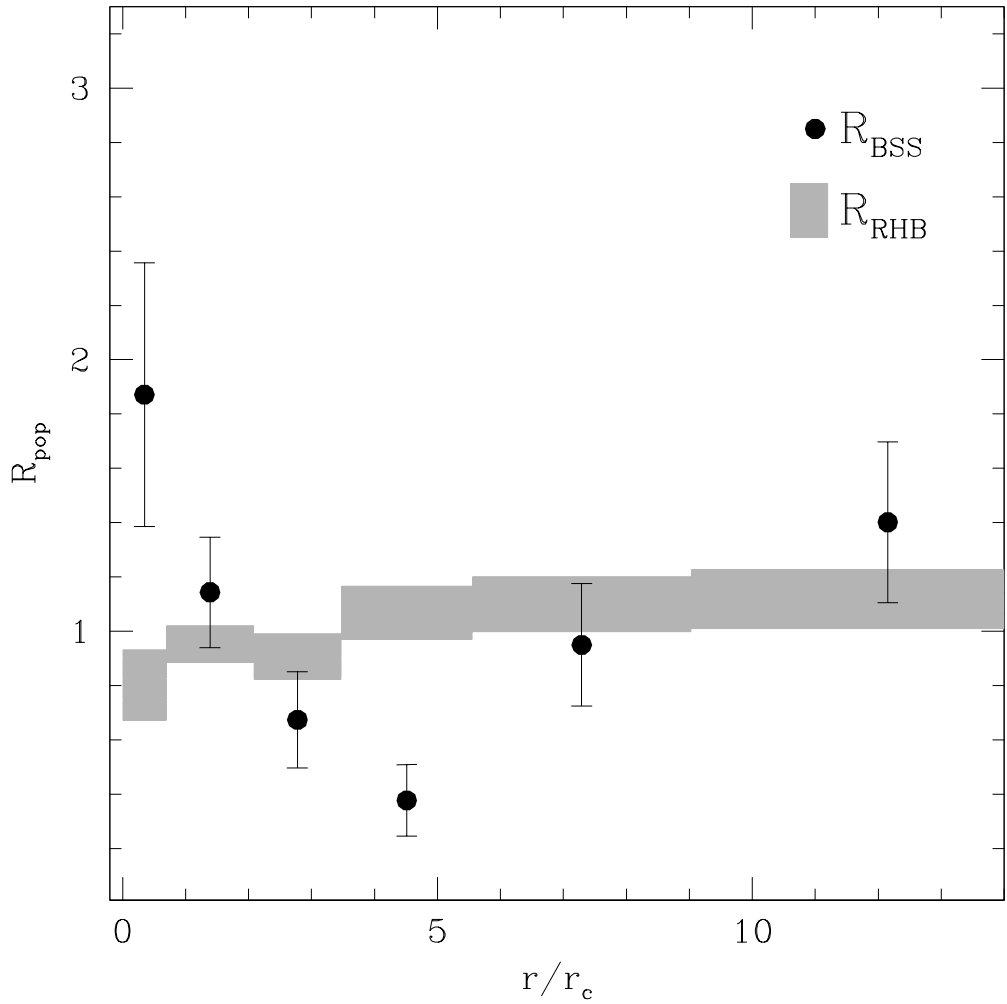}
\caption{Radial distribution of the BSS and HB double normalized ratios, as
defined in equation (\ref{eq:Rpop}), plotted as a function of the radial
coordinate expressed in units of the core radius. $R_{\rm RHB}$ (with the size
of the rectangles corresponding to the error bars computed as described in
Sabbi et al. 2004) is almost constant around unity over the entire cluster
extension, as expected for any normal, non-segregated cluster
population. Instead, the radial trend of $R_{\rm BSS}$ (dots with error bars)
is bimodal: highly peaked in the center (more than a factor of $\sim$ 2
higher than $R_ {RHB}$), decreasing at intermediate radii, and rising again
outward.}
\label{fig:Rpop}
\end{center}
\end{figure}

As expected from stellar evolution theory (Renzini \& Fusi Pecci
1988), the radial trend of $R_{\rm RHB}$ is essentially constant with
a value close to unity. On the contrary, the BSS radial distribution
is very different and is clearly bimodal. As shown in
Figure~\ref{fig:Rpop}, $R_{\rm BSS}$ reaches a value of almost two at
the center, while $R_{\rm RHB}$ has no central peak. $R_{\rm BSS}$
decreases to a minimun near $r=5\,r_c$, and rises again near
$r=11\,r_c$.

\section{DISCUSSION}
\label{sec:discus}
\begin{figure}[!p]
\begin{center}
\includegraphics[scale=0.7]{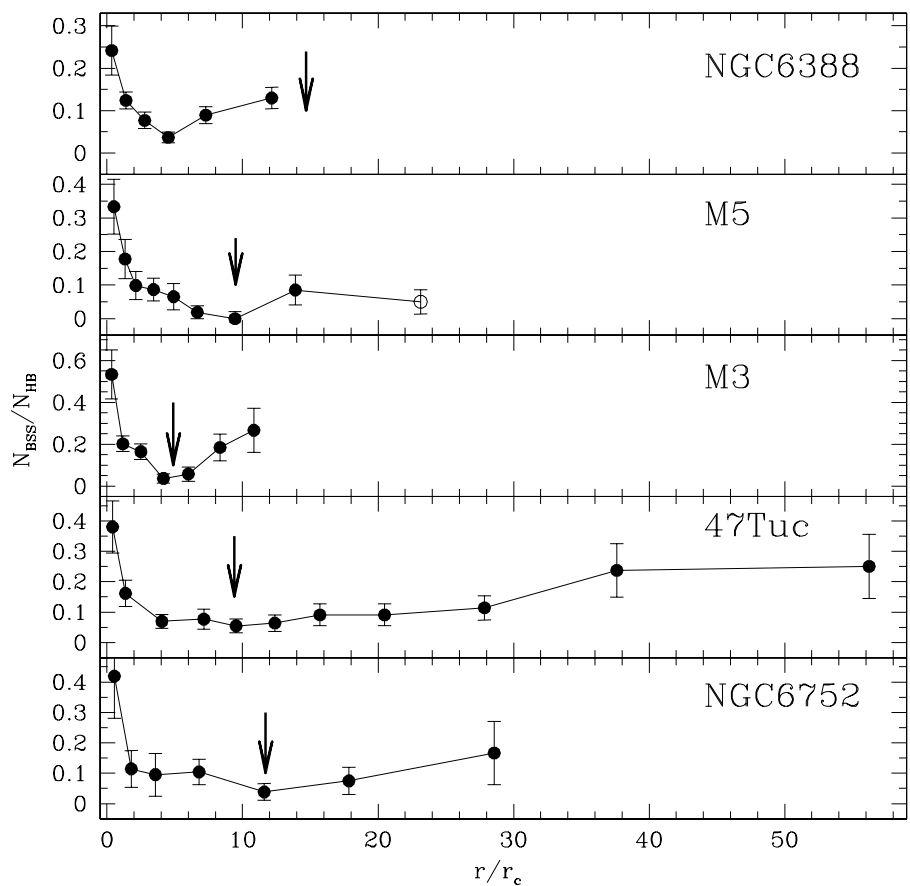}
\caption{Radial distributions of the specific frequency $N_{\rm BSS}/N_{\rm
HB}$, as observed for NGC~6388 and for other four clusters \citep[see
references in][]{lan07_M5}. The arrows mark the position of the estimated
radius of avoidance (see Sect.~\ref{sec:discus}). This well corresponds to
the position of the observed minimum of distributions, but in the case of
NGC~6388.}
\label{fig:confro}
\end{center}
\end{figure}

In Figure~\ref{fig:confro} we have compared the radial distribution of the
ratio between the BSS and HB number counts computed for NGC~6388, with that
obtained for other GCs showing a bimodal distribution \citep[see for
example][]{lan07_M5}.  The position of the observed minimum in NGC~6388
resembles that of M3, but NGC~6388 has a core radius $\sim$ 3 times smaller.
In physical units its minimum is closer to the cluster center than in any
previously observed cluster.  By equating the dynamical friction timescale
\citep[$t_{df}\propto \sigma^3 /\rho$;][see also Mapelli et al. 2006]{BT} to
the cluster age (assumed to be $t=12\,$Gyr), one can estimate the value of
the radius of avoidance ($r_{\rm avoid}$). This is defined as the radius
within which all the stars of $\sim 1.2\,M_\odot$ (the expected average mass
for BSS) have already sunk to the core because of dynamical friction effects.
As shown in Fig.~\ref{fig:confro}, the position of $r_{\rm avoid}$ well
corresponds to that of the observed minimum for all the clusters studied to
date in a similar way.  For NGC~6388, by adopting $\sigma_0= 18.9 \,{\rm
km\,s^{-1}}$ and $n_0= 10^6 \,{\rm \,stars\,pc^{-3}}$ as central velocity
dispersion and stellar density \citep{pry93}, and by assuming the cluster
structural parameters derived from the best-fit King model by \citet[][see
Table~\ref{tab:param}]{lan07bh}, we have obtained $r_{\rm avoid}\simeq
15\,r_c$.  This is about 3 times larger than the location of observed
minimum, thus representing the first case where the two distances do not
coincide.  This result is quite puzzling and somehow suggests that NGC6388
appears ``dynamically younger'' than expected on the basis of its structural
properties.  In fact, our observations suggest that the dynamical friction in
this cluster has been effective in segregating BSS (and similar massive
objects) out to only 4--5 $r_c$, whereas the theoretical expectation
indicates that, within $15\,r_c$, all stars with the mass of BSS or their
binary progenitors should have already sunk to the center.  Note that
significantly larger (by a factor of 2) velocity dispersion, or lower (by a
factor of 7) central density would be necessary to reconcile the expected and
the observed minima.  Why is dynamical friction less efficient in this
cluster?  Could the presence of an IMBH in the cluster center be important?
As discussed in Lanzoni et al (2007), the radius of influence of a $5\times
10^3\, M_{\odot}$ BH at the center of NGC~6388 is only 0.07\,pc or
0.15\,$r_c$, so it is not obvious how it might affect cluster evolution at
5--15\,$r_c$. Perhaps the BSS that we observe at $r \gg r_c$ are stars which
have ``visited'' the central region and have been put on highly eccentric
orbits by the interaction with the BH. However this effect would probably
fill-in the BSS avoidance region in the projected radial distribution.

Could the BHk population, the suspected presence of a central IMBH, the
anomalous position of the BSS radius of avoidance be connected to each other?
While it is tempting to think that such a connection exists, we are at a loss
to what that connection actually is.

\acknowledgements 
This research was supported by contract ASI-INAF I/023/05/0, PRIN-INAF2006
and by the Ministero dell'Istruzione, dell'Universit\'a e della Ricerca, and
it is also part of the {\it Progetti Strategici di Ateneo 2006} granted by
the University of Bologna. ED is supported by ASI. RTR is partially supported
by STScI grant GO-10524.  This research has made use of the ESO/ST-ECF
Science Archive facility which is a joint collaboration of the European
Southern Observatory and the Space Telescope - European Coordinating
Facility.

\begin{deluxetable}{lp{7cm}l}

\tablecaption{Papers in This Series}
\startdata \\
\hline 
\hline
NAME  &  ARGUMENT &   REFERENCES\\
\hline
M3   &  First detection of a BSS bimodal radial distribution & Ferraro et al. 1993,1997\\
M80 & BSS population in the core & Ferraro et al. 1999\\
47Tuc &  BSS radial distribution & Ferraro et al. 2003\\
NGC288 & BSS population in the core & Bellazzini et al. 2002\\
47Tuc & BSS dynamical modeling and introduction of zone
       of avoidance concept & Mapelli et al. 2004\\
NGC6752 & BSS radial distribution & Sabbi et al. 2004\\
$\omega$ Cen & The first and so far only flat BSS radial distribution & Ferraro et al. 2006a\\
47Tuc & BSS chemical signatures & Ferraro et al. 2006b \\
NGC6266 & BSS population in the core & Beccari et al. 2006a\\
  --   & BSS dynamical modeling of four GCs &  Mapelli et al. 2006\\
M5     & BSS obs and theoretical studies & Lanzoni et al. 2007a\\
NGC~1904  & BSS obs and theoretical studies & Lanzoni et al. 2007b\\
M55 & BSS radial distribution & Lanzoni et al. 2007d\\
\hline
\enddata
\label{tab:papers}
\end{deluxetable}

\begin{deluxetable}{lll}
\tablecolumns{3}
\tablewidth{0pc}
\tablecaption{Adopted cluster parameters}
\startdata
\hline 
\hline
\\
Center of Gravity & $\alpha_{\rm J2000} = 17^{\rm h}\, 36^{\rm m}\, 17^{\rm s}.23$ &
$\delta_{J2000} = -44\arcdeg\,44\arcmin\, 07\farcs1$  \\
Concentration ($c$) & 1.8 & \\
Core Radius ($r_{c}$) & 7\farcs2 & 0.46\,pc \\
Tidal radius ($r_{t}$) & 454\arcsec & 29\,pc \\
Distance  & $(m-M)_0=15.60$ & 13.2\,kpc \\
Reddening & $E(B-V)=0.32$ & \\
\hline
\enddata
\label{tab:param}
\end{deluxetable}

\begin{deluxetable}{lccccccc}
\footnotesize
\tablewidth{15cm}
\tablecaption{The BSS population of NGC~6388}
\startdata \\
\hline \hline
Name   &    RA       &   DEC     &$m_{255}$& U   &   B   &   V & \\
       & [degree]    &  [degree] &         &     &       &     & \\
\hline
\\
BSS-1 & 264.0713276 & -44.7367013 & 19.586 & 19.232 & 19.995 &  ---    \\
BSS-2 & 264.0744196 & -44.7337510 & 19.883 & 19.252 & 18.321 & 17.876  \\
BSS-3 & 264.0662077 & -44.7330923 & 20.279 & 19.453 & 18.507 & 17.842 \\
BSS-4 & 264.0684376 & -44.7363620 & 20.298 & 19.463 & 18.457 & 17.709 \\
.......  & & & & & & \\ 
\hline
\enddata
\tablecomments{The complete table is available in electronic form.}
\label{tab:bss}
\end{deluxetable}

\begin{deluxetable}{rrrrrrrrrrrrc}
\tablecolumns{12}
\tablewidth{0pc}
\tablecaption{Number Counts of BSS and HB sub-population Stars}
\startdata
\hline
\hline
\\
$r_i$ &  $r_e$ & \multicolumn{2}{c}{$N_{\rm BSS}$} &
\multicolumn{2}{c}{$N_{\rm RHB}$} & 
\multicolumn{2}{c}{$N_{\rm BHB}$} &
\multicolumn{2}{c}{$N_{\rm EHB}$} & 
\multicolumn{2}{c}{$N_{\rm BHk}$} & 
$L^{\rm samp}/ L^{\rm samp}_{\rm tot}$\\ 
${\rm [arcsec]}$ &${\rm [arcsec]}$ & obs & bck & obs & bck & obs & bck & obs &
bck & obs & bck &  \\
\hline
    0 &  5 &  22 & 0 &  91 & 0 & 18& 0 & 1 &0 &6 &0 & 0.08 \\  
    5 &  15 & 43 & 0 & 346 & 0 & 57& 0 & 7 &0 &8 &0 & 0.26 \\	
   15 &  25 & 17 & 0 & 221 & 0 & 43& 0 & 3 &0 &5 &0 & 0.18 \\  
   25 &  40 & 10 & 1 & 247 & 1 & 54& 0 & 3 &0 &4 &0 & 0.17 \\  
   40 &  65 & 24 & 2 & 248 & 2 & 52& 1 & 8 &0 &8 &0 & 0.16 \\  
   65 & 110 & 37 & 7 & 237 & 6 & 40& 1 & 4 &0 &6 &0 & 0.15 \\
\hline
\enddata
\tablecomments{The values listed as ``obs'' correspond to the observed number
of stars counted, while those listed as ``bck'' quote the estimated number of
contaminating field stars in each radial annulus (see
Sect.~\ref{sec:contamination}).}
\label{tab:decon}
\end{deluxetable} 

\newpage

\end{document}